\documentclass[12pt]{article}

     \usepackage{graphics}

\begin{document}

 \begin{titlepage}
\begin{flushright}
WISC-TH-98-348
\end{flushright}

\vspace{0.4in}

\begin{center}

  \bf{THE SPECIFIC HEAT OF A FERMI GAS AT LOW TEMPERATURE:}
	\\ \textbf{A closer look at the T lnT behavior}
\end{center}

\vspace{0.9 in}

\begin{center}

 \bf{A. REBEI}\footnote{rebei@math.wisc.edu}   \\  
  \it {Department of Physics  \\
	 University of Wisconsin. Madison, WI 53706}\\

\vspace{0.3 in}
	\bf{ W.N.G. HITCHON}\footnote{hitchon@engr.wisc.edu}  \\
	\it{Department of Electrical and Computer Engineering   \\
            University of Wisconsin. Madison, WI 53706 }

\end{center}
\vspace{0.5 in}

\begin{abstract}

At finite temperature, a Fermi gas can have states that hold simultaneously a 
particle and a  hole with a finite probability. This gives rise to a new set
of diagrams that are absent at zero temperature. The so called `` anomalous''
diagram is just one of the new diagrams. We have already studied the contribution
of these new diagrams to the thermodynamic potential (\textit{Phys. Lett.}
A224,127 (1996)). 
Here we continue
that work and calculate their effect on the specific heat. We will also calculate
the finite temperature contribution of the ring diagrams. We conclude that the
$T lnT$ behavior due to exchange gets canceled by the new contribution of the new
diagrams and screening is not essential to solve this anomaly. 
\end{abstract}

\end{titlepage}

In this letter we calculate the specific heat of an electron gas at low temperature
and high density. This calculation is not just an improvement
over
 the Gell-Mann calculation \cite{gellmann}. It is fundamentally different. 
The Seitz formula used by Gell-Mann has been criticized in
the literature and the relevance of screening has been questioned
\cite{isihara},
\cite{kraeft}. We would like to stress  that we are not arguing that
the divergence of the inverse of the density of states 
at the Fermi level is not due to the long range
nature of the Coulomb field. Here instead we are addressing the original problem of the
$T ln T $ behavior of the specific heat that was pointed out by Bardeen 
\cite{bardeen} (see also \cite{wigner})
 if exchange is included. In fact our calculation shows that there is
no need to invoke the screening of the Coulomb potential to find a linear
function for the specific heat at low temperature. We stress that 
the solution to this problem has apparently not been clarified yet! 
Clearly finite 
temperature calculations show that the problem is related to the Fermi distribution.
In fact all of these calculations \cite{horowitz} give a finite answer for the specific 
heat with exchange included. Unfortunately, Gell-Mann in his paper did not settle
this point, even though he cited Bardeen's paper. This caused a confusion between
two different types of logarithmic behaviors. One is due to the long range behavior
of the Coulomb field while the other is due to the statistic of the 
Fermi particles. In 
this letter we show that the specific heat maintains its observed linear behavior
at low temperature without one having to worry
about screening. However as we showed in 
a previous letter \cite{rebei} we have to include a new infinite set of diagrams.  
This new set does not contribute anything at zero temperature. The diagrams are merely due
to the fact that at nonzero temperature there is a finite probability of finding
a state that is occupied by a particle and a hole. 

\begin{figure}
\resizebox{\textwidth}{!}
{\includegraphics[0in,0in][8in,10in]{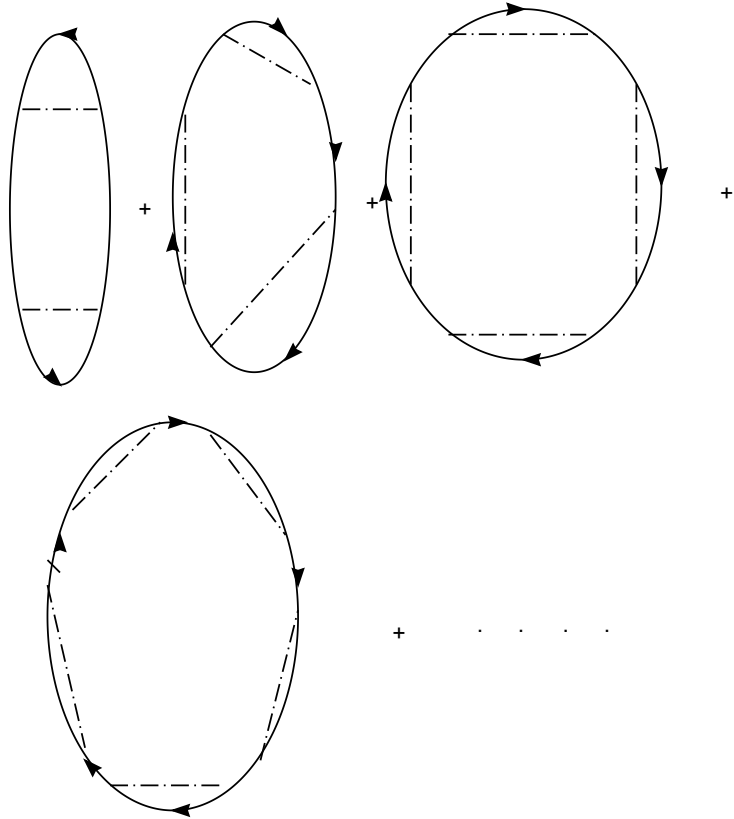}}
\caption{A new infinite set of ``anomalous'' diagrams. In this approximation,
the solid line corresponds
to the  Fermi propagator and the dashed line corresponds to the  Coulomb
field.}
\label{anomalous}
\end{figure}

We start as usual from the Hamiltonian of the system, 

\begin{eqnarray}
 H_{0} & = & \int d^{3} x \lbrack - 
\frac{1}{2} 
			\Psi_{\alpha}^{\dagger} (x)
                 \nabla^{2} \Psi_{\alpha} (x) \rbrack \nonumber  \\
       &&  + 1/2  \int d^{3}x d^{3}y \frac{1}{|\vec{x}-\vec{y}|} 
               \Psi_{\alpha}^{\dagger}(x) \Psi_{\beta}^{\dagger} (y) 
\Psi_{\beta} (y) 
               \Psi_{\alpha} (x) 
\end{eqnarray}\\
 Here, we use units such that $\hbar =m=e=1$. $\Psi (x)$ is a
 two-component
electron field. $\alpha $ and $\beta $ are spin indices, i.e., $\alpha =1$
for spin up and $\alpha =-1$ for spin down. Summation is implicit for
repeated indices. The system is constrained
by the condition
 
\begin{equation}
\int d^{3}x\Psi _{\alpha }^{\dagger }(x)\Psi _{\alpha }(x)=N
\end{equation}
where N is the electron number operator which is constant. 
 Because of this
constraint, we prefer to work  with the  Hamiltonian 

\begin{equation}
H=H_{0}-\mu N
\end{equation}
where $\mu $ is a Lagrangian multiplier.
This calculation we present is  mainly an improvement on calculations of the exchange energy and ring
diagrams made by Isihara and
coworkers \cite{isihara} and others. 
The appearance of temperature dependent logarithmic
terms in the internal energy found through various calculations are unacceptable. 
These terms will reappear
again in the specific heat and will spoil the observed linear behaviour at
low temperature. In the past these difficulties were not given close
attention. They were either hidden in an effective mass or the expansion was
not completely correct as we argue below. The history of this problem dates
back to the '30s when it was first pointed out by Bardeen \cite{bardeen} . A
preliminary treatment of this work appeared in \cite{rebei}. It was argued
that a correct approximation to the internal energy of a many-particle
system up to exchange, i.e., a Hartree-Fock approximation, will
automatically include summing an infinite set of diagrams 
 (figure~\ref{anomalous}). The first diagram in this set is known as the
 ``anomalous'' diagram \cite{kohn}. \\
\noindent Therefore
 any approximations must be made at the
level of two-particle Green's functions and not the single-particle wave
functions as was done in most past treatments. This is exactly the reason for
introducing two-point sources instead of the one-point source used in the usual
effective potential method \cite{rebei}, \cite{domincis}. We have seen that
 this enabled us to write an expression
for the energy in terms of the true two-particle Green's functions of the system. 
Hence for the two-body
Coulomb potential a Hartree-Fock approximation is equivalent to writing the
two-particle Green's function in terms of the one-particle Green's function as
follows:

\begin{equation}
G(x_{1},x_{2};x_{3},x_{4})=G(x_{1},x_{3})G(x_{2},x_{4})-G(x_{1},x_{4})G(x_{2},x_{3})
\end{equation}\\
This approximation is equivalent, as it is well known, to the usual Hartree-Fock 
approximation of
the energy to order $e^{2}$ at zero temperature. At low temperature there is
another contribution to order $e^{2}$ that comes from summing an infinite
set of diagrams that were never treated before. The ring diagrams treated in 
\cite{gellmann} are of order $e^{4}$ and therefore the new set can not be
ignored in any finite temperature treatment. The effective action method can
be easily extended to the non-zero temperature case. In the 
presence of external sources, the grand partition
function $Z$ of the system is given in this case by the following
expression:

\begin{eqnarray}
Z=\exp ( -\beta \Omega )=Tr \exp \left( -\beta \left [
\left ( \rm H - \mu \rm N \right) + j \varphi + \frac{1}{2} \varphi \rm B \varphi +
\psi^{ \dag} \rm Q \psi \right] \right) \: .
\end{eqnarray}\\
The trace is over space indices, internal indices and ``imaginary time''.
The imaginary time interval is $[0,\beta ]$. The thermodynamic potential $%
{\Omega }$ should be considered as a functional of the sources $j(x,\tau )$, $%
B(x,\tau ;x^{\prime },\tau ^{\prime })$ and $Q(x,\tau ;x^{\prime },\tau
^{\prime })$. The other parameters in the system are the variables $\mu ,T$
and $V$.
Now we define new variables by the following functional relations: 

\begin{eqnarray}{\frac{\delta  \Omega}{\delta j}}=\varphi_{c} \: ,\end{eqnarray}
\begin{eqnarray}{\frac{\delta \Omega}{\delta B}}=\Delta +\frac{1}{2}\varphi_{c}\varphi_{c}
\end{eqnarray}
and
\begin{eqnarray}\frac{\delta\Omega}{\delta Q}=\rho \: .\end{eqnarray}\\
The above equations are evaluated at $j=B=Q=0$. To get the free energy $
F[\rho ,\varphi _{c},\Delta ]$ we make the following triple
Legendre transformation
\begin{eqnarray}
 F[\rho,\varphi_{c},\Delta]-\mu \,N =  \Omega [j,B,Q] -j\varphi_{c} -\frac{1}{2}
\varphi_{c}B\varphi_{c} -\rho Q -\Delta B
\end{eqnarray}
To calculate ${F}$ perturbatively, we expand the exponent around the
classical Hartree potential in a neutral background. The coefficients of
this expansion are expressible in terms of Feynman diagrams. 
We keep only the first two terms of the expansion. The
equations of motion of the variables $\rho $, $\varphi _{c}$  and $\Delta $
are obtained by the standard way. The variables $\varphi _{c}$ and $\Delta $
can be expressed as a series expansion in $e^{2}$ with coefficients
dependent only on $\rho $.  Hence within the above approximation, the expression
for the thermodynamic potential has the following form:

\begin{eqnarray}
\Omega=\Omega_{0}+\Omega_{ex}+
\frac{1}{2}Tr\{\ln\left[\delta(x-y) +e^{2}\Re(x-y)\right]-e^{2}\Re(x-y)\} \nonumber  \\
-Tr\{\ln\left[\delta(x-y)-e^{2}\Im(x-y)\right]+e^{2}\Im(x-y)\}
\end{eqnarray}
where the functions $\Re(x-y)$ and $\Im(x-y)$ are given by
\begin{eqnarray}\Re (x-y) = \int_{0}^{\beta}d\tau\int d^{3}z A(y-z)\rho (x-y)\rho (z-x)
\end{eqnarray}
and
\begin{eqnarray} \Im (x-y)=\int_{0}^{\beta}d\tau \int d^{3}z A(x-z)\rho(z-x)\rho(y-z).
\end{eqnarray}
The term $\Omega_{0}$ is the free contribution, $\Omega
_{ex}$ is the usual first order exchange term, the third term represents the
usual ring diagrams and the last term represents a new series of diagrams
shown in figure~\ref{anomalous}. The function $A(x-y)$ is the bare Coulomb potential and $%
\rho (x-y)$ is the free one-particle Green's function at finite temperature.
The signs in front of the two $Tr\,\ln$-terms are determined by integrating over
Bose fields and Fermi fields, respectively. We have shown that the new
diagrams provide a $T\,\ln T$ term with the right weight to cancel
 the $T lnT$ term that
appears in the exchange term. This term is beyond the reach of 
all previous treatments including those 
  of Gell-Mann and Isihara. Hence the unphysical logarithmic behavior
of the exchange energy gets cancelled by a consistent perturbative treatment
of the many-body problem. The other important point we learned from this
calculation is that this logarithmic term is not an immediate byproduct of
the singular behavior of the Coulomb potential since these new diagrams are
not singular like the ring diagrams. The new contribution at high density
is given by 
\begin{eqnarray}
\Gamma=\frac{e^{2}V}{12\pi\eta^{2}} k_{f}^4(\zeta + \frac{1}{2}\ln\eta).
\end{eqnarray} 

\noindent The ring diagrams contribute a 
temperature dependent term of order $e^{4}$ \cite
{isihara} as we will show next. The ring 
diagrams are a manifestation of the long range behavior
of the Coulomb field. In fact, in our calculation, the ring diagrams are
obtained by integrating over the Hartree field while the new terms are
obtained by integrating over the Fermi fields. We do not think that this
point has been recognized in the literature. Obviously, these new diagrams
have a nonzero contribution to the specific heat and must not be left out in
any accurate calculation. Their contribution appears when we try to
calculate $\mu $ in terms of the Fermi energy of the free electron gas. The
 fact that these new terms have an opposite contribution
to the exchange energy can be easily deduced if we
 refer back to \cite{rebei}. In fact, 
our method
shows that for a many-body problem with a two-body potential, isolating
the exchange energy is really an artificial effect. Reference \cite{rebei1}
 shows that the exchange
 energy
term really comes from the ring diagrams by integrating over the Hartree potential. 
However, 
the new terms came by integrating the Fermi field degrees of freedom 
which have a statistical
origin.\\ 

Now we continue evaluating the contribution
of the ring diagrams at finite temperature. This contribution is given
by 

\begin{eqnarray}
R\; = \frac {1}{2} Tr\{ln\lbrack\delta(x-y)+e^2\Re(x-y)\rbrack
-e^2\Re(x-y)\}.
\end{eqnarray}\\
In momentum representation, this is given by

\begin{eqnarray}
R\; = \frac{V}{2\beta}\, \sum_{n=-\infty}^{n=\infty} \frac{4\pi}{(2\pi)^{3}}
\int_{0}^{\infty}\, p^{2}\,dp\; \lbrack \ln \lbrace 1 + e^{2} \Re (p,\omega_{n})
\rbrace -e^{2} \Re (p,\omega_{n}) \rbrack
\end{eqnarray}\\
where p represents momentum and $\omega_{n}=\frac{(2n+1)\pi}{\beta}$, $n=0,\pm 1,
\pm 2, \ldots $. The function $\Re (p,\omega_{n})$ is given by:

\begin{eqnarray}
\Re (p,\omega_{n})=\frac{2}{(2\pi)^{3}p^{2}}\! \int \!d\vec{q} 
\frac{\frac{1}{2}(\vec{p}+
\vec{q})^{2} -\frac{1}{2}\vec{q}^{2}} {\omega_{n}^{2} + 
( \frac{1}{2}(\vec{p}+\vec{q})^{2}
- \frac{1}{2} \vec{q}^{2})^{2}} (f(\vec{q})-f(\vec{p}+\vec{q})) .
\end{eqnarray}

\noindent The function $f(\vec{q})$ is the Fermi-Dirac function,

\begin{eqnarray}
f(\vec{q})\;=\; \frac{1}{ 1 + \exp \lbrack \beta (q^{2}/2 - \mu)\rbrack}  .
\end{eqnarray}

\noindent To evaluate R we break  
up $\Re(p,\omega_{n})$ in two pieces. We write
\begin{eqnarray}
\Re(p,\omega_{n})= F_{1}(p,\omega_{n})+F_{2}(p,\omega_{n})
\end{eqnarray}
such that
\begin{eqnarray}
F_{1}(p,\omega_{n})=\frac{2}{(2\pi)^{3}p^{2}}\int d\vec{q} \frac{E_{\vec{p}+\vec{q}}
-E_{\vec{q}}}{(E_{\vec{p}+\vec{q}}-E_{\vec{q}})^{2}+\omega_{n}^{2}} f(q)
\end{eqnarray}\\
where $E_{\vec{q}}= {\frac{ \vec{q}^{2}}{2}}$. The other function $F_{2}
(p,\omega_{n})$
is similarly defined but with $ f(\vec{q})$ replaced by
$ f(\vec{p}+\vec{q})$.\\
 We start by evaluating the function $ F_{1}(p,\omega_{n})$.
Taking $\vec{p}$ along the z-axis and integrating the angular coordinates, we get

\begin{eqnarray}
F_{1}(p,\omega_{n})\!=\! \frac{1}{(2\pi)^{2}p^{3}} \int_{0}^{\infty}\!q\, dq\;
\ln \lbrack \frac{\omega_{n}^{2}+(pq + p^{2}/2)^{2}}{\omega_{n}^{2}+
(pq - p^{2}/2)^{2}} \rbrack \frac{1}{1 + \exp \lbrack \beta ( q^{2}/2 - \mu) \rbrack}.
\nonumber \\
\end{eqnarray}\\
Now we rewrite this in the following form:

\begin{eqnarray*}
F_{1}(p,\omega_{n})\;= \; \frac{1}{(2\pi)^{2}p^{3}} \int_{0}^{k_{F}} q\,dq
\ln \lbrack \;\frac{\omega_{n}^{2}+(pq+p^{2}/2)^{2}}{\omega_{n}^{2}+(pq-p^{2}/2)^{2}}
  \; \rbrack   
\end{eqnarray*}
\begin{eqnarray*}
- \frac{ 1} {(2\pi)^{2}p^{3}} \int_{0}^{k_{F}} q\, dq
\ln \lbrack \frac{\omega_{n}^{2}+(pq+p^{2}/2)^{2}}{\omega_{n}^{2}+(pq-p^{2}/2)^{2}}
\rbrace \frac{1} {1+\exp\lbrace -\beta(q^{2}/2 -\mu) \rbrack} 
\end{eqnarray*}
\begin{eqnarray}
+ \frac{ 1 }{ (2\pi)^{2}p^{3}} \int_{k_{F}}^{\infty} q\, dq
\ln \lbrack \frac{\omega_{n}^{2}+(pq+p^{2}/2)^{2}}{\omega_{n}^{2}+(pq-p^{2}/2)^{2}}\rbrack
 \frac{1}{ 1+\exp \lbrace \beta (q^{2}/2 - \mu)\rbrace } \: .
\end{eqnarray}\\
In the limit when $\beta \rightarrow \infty$, the first integral that we isolated 
becomes simply part of the zero temperature contribution to $R$. We denote this part
by $F_{1,0}(p,\omega_{n})$. Next we set $z=\beta (q^{2}/2 -\mu)$ and integrate over 
z instead of q. Since we are interested in the low temperature limit, we get to 
leading order in $1/\beta$ that

\begin{eqnarray}
F_{1}(p,\omega_{n})&=&F_{1,0}(p,\omega_{n})+\frac{\pi}{(2\pi)^{3}p^{3}}\frac{1}{\beta}
\int_{0}^{\infty} \frac{dz}{1 +\exp z}\lbrace \ln \lbrack \frac{ \omega_{n}^{2}+
(p^{2}/2 +pk_{F}(1+ \frac{z}{2\eta}))^{2}}{\omega_{n}^{2}+(p^{2}/2-pk_{F}(1+
\frac{z}{2\eta}))^{2}}\rbrack \nonumber  \\
&&- \ln \lbrack \frac{\omega_{n}^{2}+(p^{2}/2+pk_{F}(1-\frac{z}{2\eta}))^{2}}
{\omega_{n}^{2}+(p^{2}/2-pk_{F}(1-\frac{z}{2\eta}))^{2}}\rbrack \rbrace \: .
\end{eqnarray}

\noindent Again this last integral over z can be separated into four integrals that
can all be evaluated  in the same way by elementary methods. To leading order in
$\beta$, we find

\begin{eqnarray}
F_{1}(p,\omega_{n})\!=\!F_{1,0}(p,\omega_{n})\!+\!\frac{1}{(2\pi)^{2}pk_{F}}
\frac{\pi^{2}}{12\beta^{2}} ( \frac{ \frac{p}{2}+k_{F}}
{ \omega_{n}^{2} +(p^{2}/2 +pk_{F})^{2}}\! +\! \frac{ \frac{p}{2}- k_{F}}
{\omega_{n}^{2} + (p^{2}/2 - pk_{F})^{2}}) \nonumber \\
\end{eqnarray}

\noindent The function $F_{2}(p,\omega_{n})$ is 
easily seen to be equal to $F_{1}(p,\omega_{n})
$, hence to this approximation the function $\Re(p,\omega_{n})$ is equal to
\begin{eqnarray}
\Re(p,\omega_{n})\, =\, 2 F_{1,0}(p,\omega_{n}) + 2 F_{1}(p,\omega_{n}) \:.
\end{eqnarray}
Putting his back into $R$, and expanding the logarithmic function in powers of 
${\frac{1}{\eta}}$, we get

\begin{eqnarray}
R = \frac{V}{2\beta} \frac{4\pi}{(2\pi)^{3}} \sum_{n=-\infty}^{n=\infty}
\int_{0}^{\infty}p^{2}\, dp \, \ln \lbrace 1+2e^{2}F_{1,0}(p,\omega_{n})
\rbrace - 2e^{2} F_{1,0}(p,\omega_{n})  \nonumber \\
- \frac{V}{2\beta} \frac{4\pi e^{2}}{(2\pi)^{3}} \sum_{n=-\infty}^{n=\infty}\frac{1}
{\beta^{2}} \int_{0}^{\infty}p^{2} \, dp \frac{2 e^{2} F_{1,0}(p,\omega_{n})}
{1 + 2 e^{2} F_{1,0}(p,\omega_{n})}\, A(p,\omega_{n})
\end{eqnarray}

\noindent where we have defined $A(p,\omega_{n})$ to be

\begin{eqnarray}
A(p,\omega_{n})=\frac{1}{24pk_{F}} \lbrace\frac{\frac{p}{2} +k_{F}}{\omega_{n}^{2}+
(p^{2}/2 +pk_{F})^{2}} + \frac{p/2 - k_{F}}{\omega_{n}^{2}+(p^{2}/2-pk_{F})^{2}}
\rbrace \: .
\end{eqnarray}

\noindent The first integral in $R$ will be 
denoted by $R_{0}$ and the second one by $R_{\beta
}$. So far our approximation works for any density. Now we notice that $F_{1,0}(p,
\omega_{n})$ is proportional to the inverse density. Hence for high densities the 
second integral can be expanded in powers of inverse densities. To lowest order,
the second part can be easily evaluated. Letting ${x=\frac{p}{k_{F}}}$ and
${y_{n}}={ \frac{\omega_{n}}{ k_{F}^{2}}}$, we have

\begin{eqnarray*}
R_{\beta}\!=\! -\frac{e^{4}}{12(2\pi)^{4}k_{F}^{3}\beta^{2}}\frac{V}{\beta}\!
\sum_{n=-\infty}^{n=\infty}\! \int_{0}^{\infty}\! \frac{dx}{x^{2}}
\left( \frac{x/2 +1}{y_{n}^{2}+x^{2}(x/2 +1)^{2}} + \frac{x/2 - 1}
{y_{n}^{2} + x^{2}(x/2 - 1)^{2}} \right)
\end{eqnarray*}
\begin{eqnarray}
 \lbrace x - y_{n} \arctan \lbrace
\frac{x(x/2 -1)}{y_{n}} \rbrace -y_{n} \arctan 
\lbrace \frac{x(x/2 -1)}{y_{n}} \rbrace  \nonumber  \\
 + 1/2 \ln \lbrack \frac{y_{n}^{2}+x^{2}(x/2 +1)^{2}}{y_{n}^{2}
 +x^{2}(x/2 -1)^{2}}\rbrack + \frac{1}{8x^{2}}(4y_{n}^{2}-x^{4}) \ln \lbrack
\frac{y_{n}^{2}+x^{2}(x/2 +1)^{2}}{y_{n}^{2}+ 
x^{2}(x/2 -1)^{2}}\rbrack \,\, \rbrace.
\end{eqnarray}

\noindent In the limit that $\beta \rightarrow \infty$, the 
summation over n can be changed
to an integral over y with measure $dy=\frac{2\pi}{\beta k_{F}^{2}}$. This integral
has been evaluated numerically. It turns out that integrating over the variable
${\frac{y}{x}}$ instead of $y$ makes the integration more suitable from
the numerical point of view. The numerical value of the integral

\begin{eqnarray}
I \,=\, \int_{0}^{\infty} \frac{dx}{x^{2}} \int_{0}^{\infty} dz \,K(x,z)\,P(x,z)
\end{eqnarray}\\
is approximately equal to $-22.94$. In the above the functions $K(x,z)$ and
$P(x,z)$ are given by

\begin{eqnarray}
K(x,z)= \frac{1+ x/2}{z^{2}+(1+x/2)^{2}} - \frac{1- x/2}{z^{2}+(1- x/2)^{2}}
\end{eqnarray}\\
and
\begin{eqnarray}
P(x,z)=1-z\arctan\left( \frac{1+x/2}{z} \right)-z\arctan \left(\frac{1-x/2}{z}
\right)\nonumber \\
 + \frac{1}{2x} \left(1-x^{2}/4 +z^{2}\right) \ln \lbrack
\frac{z^{2}+(1+/2)^{2}}{z^{2}+(1-x/2)^{2}}\rbrack \: .
\end{eqnarray}\\
Therefore the finite temperature contribution of the ring diagrams to leading
order in ${\frac{1}{\eta^{2}}}$ and for high density is given by
\begin{eqnarray}
R\,=\, N  (0.0622 \ln r_{s} -0.142 ) +0.0181\, N(T)\mu
(\frac{\alpha r_{s}(T)}{\eta})^{2}
\end{eqnarray}\\
Here N is the total number of particles, which is a constant. All the other
variables are defined as follows:

\begin{eqnarray}
\alpha = (\frac{4}{9\pi})^{1/3},  r_{s}(T)= \frac{me^{2}}{\alpha {\hbar^{2}} k_{F}},
&& \, a_{0}=\frac{\hbar^{2}}{me^{2}}, \nonumber \\
N(T)=\frac{V}{\frac{4\pi}{3}r_{s}(T)^{3}a_{0}^{3}}  .&& 
\end{eqnarray}\\
Note that here we did not resort to any artificial cutoff of the
momentum in our calculation as was done by Gell-Mann \cite{gellmann} or Isihara
\cite{isihara}. The first author assumed $T=0$, while the second author
 and coworkers carried out the integral up to
$p \leq 2k_{F}$. In fact momenta larger than $2k_{F}$ 
contribute about 5 percent to the
answer.

Our calculation of the specific heat starts by finding
 an expression for $\mu$ in terms of the Fermi energy at zero
temperature. In other words, we expand $\mu$ in powers of $\frac{1}{\beta^{2}}$. We need to do 
this before any differentiation with respect to temperature, to get the free 
energy function which is by  definition a function of T, V, and N. Here we will
need to carry our expansion only to first order in $\frac{1}{\beta^{2}}$. This 
will at least give us the first nontrivial contribution beyond the ideal one. Hence
in this case it will be enough to use for $\mu$ the equivalent free electron
expression, 
that is we take,

\begin{eqnarray}
\mu \, = \, \mu_{0}(1-\frac{\pi^{2}}{12 \eta_{0}^{2}}) \: .
\end{eqnarray}
\medskip
This implies that

\begin{eqnarray}
N(T) =  N( 1 - \frac{\pi^{2}}{8\eta_{0}^{2}}) \: .
\end{eqnarray}\\
However for $ r_{s}(T)$, it is enough to take it equal to the corresponding
zero temperature value $r_{s}$. The calculations that lead to the determination
of the temperature dependent terms in the free energy are straightforward. The ring
diagrams will contribute a term $R_{T}$, which in Rydbergs is given by

\begin{eqnarray}
R_{T}= \frac{N}{\eta_{0}^{2}} \lbrack\: (1 - \frac{\pi^{2}}{8})(0.0622 \ln r_{s}
 -0.142) + \frac{22.90}{64(2\pi)^{2}} \:\rbrack \: .
\end{eqnarray}
\medskip
Similarly, the ``anomalous'' diagrams contribute a term of the form
\begin{eqnarray}
A_{T}=-\frac{3\pi \,N}{2\eta_{0}^{2} \alpha  r_{s}}(\zeta +\frac{1}{2}\ln
\eta_{0}) \: .
\end{eqnarray}
The constant $\zeta$ is approximately equal to $8.0$. Besides these, we have
the usual contribution from a kinetic part and an exchange part. They are, respectively,
 given by

\begin{eqnarray}
K_{T}= \frac{2.21 N}{r_{s}^{2}} \frac{5\pi^{2}}{\eta_{0}^{2}}
\end{eqnarray}\\
and

\begin{eqnarray}
E_{T}^{exch}=\frac{0.916 N}{r_{s}}\frac{\pi^{2}}{3\eta_{0}^{2}}
( 1 + \frac{3 \it{a}}{2\pi^{2}} +\frac{1}{2} \ln \eta_{0}) \: .
\end{eqnarray}\\
The constant $a$ has different values in the literature. For example Isihara and
coworkers give it a value of approximately $-\frac{1}{2}$. Our calculation gives it
a value of almost $+2.0$. Both values give  similar answers to the 
specific heat. It will be seen below that the contribution from 
the new diagrams 
is much more important and hence any discrepancies in the values of $\it{a}$ will
 hardly matter.\\
 Remembering that $F(T,V,N)=\Omega +\mu N$, we can easily find
the contribution of all of the above terms to the temperature dependent term,
 $F_{T}$, of the free energy $F$ using standard techniques \cite{fetter}.
 Finally, we can now get an estimate for the
specific heat ,

\begin{eqnarray}
C_{V}\, = \, -T \frac{\partial^{2} F}{\partial T^{2}}
\end{eqnarray}
such that
\begin{eqnarray}
F(T,V,N) \,=\,F_{0}(T,V,N)+ \frac{k^{2} T^{2}}{\mu_{0}^{2}}
\left( -\frac{20.19}{r_{s}} -0.077\ln r_{s}
+ 0.184 \right)
\end{eqnarray}\\
and $F_{0}$ is the free energy of a free electron gas.
Therefore, if $C_{V}^{0}$ is the specific heat for the ideal Fermi gas, we get
for the specific heat at low temperatures and high densities the  
expression,

\begin{eqnarray}
\frac{C_{V}}{C_{V}^{0}} \,= \, 1.00 + 0.185 r_{s} - r_{s}^{2}
 ( 0.0017 - 0.0007 \ln r_{s}) \: .
\end{eqnarray}\\
Some comments on this expression are called for. First, without the new effect of the anomalous
diagrams, the second term would have been negative besides the $\ln T$ dependency.
Second, for sodium, which has $r_{s}= 3.93$, the ratio $C_{V}/C_{V}^{0}=1.71$ compared to
the experimental value of 1.26. This is not a bad outcome considering that our
calculation is strictly valid for only small $r_{s}$. Gell-Mann's computation gives
0.48 as an answer in this case. Going back to Eq.(22), it is easily seen that had we kept the next
term in our approximation, we could easily improve our answer. Besides the numerical answer, it is important
to realize that screening has little effect on the specific heat and need not be
included to get a finite and reasonable answer for the specific heat. Exchange 
energy is essentially of statistical origin. At nonzero temperature a point in
phase space can simultaneously hold a particle and a hole and hence the statistics
of holes becomes important as well and  it is manifested  in the appearance of this 
new type of diagrams. The name ``anomalous'' is therefore misleading. In fact, they
should be called particle-hole exchange diagrams. This name shows that this kind
of exchange is not allowed at zero temperature.

We end this letter by stressing the fact this expansion is not an expansion
in $e^{2}$, and that what this method considers exchange 
is not the usual meaning of the word \cite{rebei1}.
 The usual first order exchange, the ring 
diagrams and the new particle-hole exchange diagrams are all part of the first
correction term to the energy of an interacting Fermi gas. The second order exchange diagram and the ring diagrams with exchange appear in the
next term of the expansion. The third term in this expansion contains among other 
things the ladder diagrams. These diagrams are important for 
metal densities. We hope 
to include all these effects in another communication. 

Finally, we would like to mention that this
 expansion with two-point functions was also recently used in the 
study of the anharmonic oscillator at low temperature \cite{anna}. Here
it was shown that effective actions with two-point functions give better
results than effective actions with one-point functions.

\newpage

\newpage

\begin{thebibliography}{99}



\bibitem{gellmann}M. Gell-Mann, \textit{Phys.Rev.}\textbf{106 }, 369 (1957) 

\bibitem{isihara}  A.Isihara, \textit{Condensed Matter Physics }. Oxford,1991.
 A. Isihara and D. Y. Kojima, \textit{Physica }\textbf{77, }469 (1974). 
A. Isihara and D. Y. Kojima, \textit{Z.Phys. }\textbf{21, }33 (1975).
  A.Isihara and D.Y.Kojima, \textit{Z.Physik}\textbf{\ 21 }, 33 (1975)




\bibitem{kraeft} W. D. Kraeft and W. Stolzmann, \textit{Phys. Lett.}\textbf{A 56}, 41 (1976)

\bibitem{bardeen}  J. Bardeen, \textit{Phys. Rev. }\textbf{50}, 1098 (1936)
\bibitem{wigner} E. Wigner, \textit{Phys. Rev. }\textbf{46}, 1002 (1934); \textit{
Trans. Faraday Soc. } \textbf{34}, 678 (1938)


\bibitem{horowitz}  B.Horovitz and R.Thieberger, \textit{Physica }\textbf{71 }, 99 (1974). I. Yokota, \textit{J. Phys. Soc. Japan} \textbf{4}, 82 (1949). A. B. Lidiard,\textit{Proc. Phys. Soc.}\textbf{A64}, 814 (1951). E. Wohlfarth, \textit{ Phil. Mag.}\textbf{41}, 534 (1950)


\bibitem{rebei}  A. Rebei and W.N.G. Hitchon, \textit{Phys. Lett. }\textbf{A224}, 127 (1996)



\bibitem{kohn}  K. Kohn and J.M. Luttinger, \textit{Phys. Rev. }\textbf{118}, 41 (1960)

\bibitem{domincis}  C.de Dominicis and P.C.Martin, \textit{J.Math.Phys.}
\textbf{5}, 14 (1964)

\bibitem{rebei1} A. Rebei and W. N. G. Hitchon, in preparation



\bibitem{fetter} A. L. Fetter and J. D. Walecka \textit{ Quantum Theory of Many-Particle Systems }. McGraw-Hill, 1971.

\bibitem{anna} A. Okopiska, \textit{Int. J. Mod. Phys.}\textbf{A 12}, 585 (1997),hep-th/9704038


\end{thebibliography}
\end{document}